 \documentclass[prl,twocolumn,showpacs,superscriptaddress,floatfix]{revtex4}
 \def\LineNumbers{false}	
 \def\AuthorList{true}

\usepackage{ifthen}
\usepackage{lineno}
\usepackage{slashed}
\usepackage{epsfig}
\usepackage{amsmath}
\usepackage{comment}
\usepackage{longtable}
\usepackage{subfigure}
\usepackage{color}

\newcommand{\pythia}{{\sc pythia}}

\newcommand{\ppbar}{$p\bar{p}$}

\newcommand{\bbbar}{$b\bar{b}$}
\newcommand{\fb}{fb$^{-1}$}
\newcommand{\BRhbb}{$\cal{B}$($H \rightarrow b\bar{b}$)}
\newcommand{\WHlvbb}{$WH \rightarrow \ell \nu b\bar{b}$}

\newcommand{\pt}{p_{T}}
\newcommand{\et}{E_{T}}

\newcommand{\gevcc}{GeV/$c^2$}
\newcommand{\gevc}{GeV/$c$}
\newcommand{\MET}{\mbox{$\raisebox{.3ex}{$\not\!$}E_T$}}
\newcommand{\METVEC}{\mbox{$\raisebox{.3ex}{$\not\!$}{\vec E}_T$}}

\def\BNN {BNN}

\begin{document}

\ifthenelse{\equal{\LineNumbers}{true}}{
 \pagewiselinenumbers
 \linenumbers
}

\title{Search for the standard model Higgs boson decaying to a $b\bar{b}$ pair 
in events with one charged lepton and large missing transverse energy 
using the full CDF data set}

\ifthenelse{\equal{\AuthorList}{true}}{
  \affiliation{Institute of Physics, Academia Sinica, Taipei, Taiwan 11529, Republic of China}
\affiliation{Argonne National Laboratory, Argonne, Illinois 60439, USA}
\affiliation{University of Athens, 157 71 Athens, Greece}
\affiliation{Institut de Fisica d'Altes Energies, ICREA, Universitat Autonoma de Barcelona, E-08193, Bellaterra (Barcelona), Spain}
\affiliation{Baylor University, Waco, Texas 76798, USA}
\affiliation{Istituto Nazionale di Fisica Nucleare Bologna, $^{ee}$University of Bologna, I-40127 Bologna, Italy}
\affiliation{University of California, Davis, Davis, California 95616, USA}
\affiliation{University of California, Los Angeles, Los Angeles, California 90024, USA}
\affiliation{Instituto de Fisica de Cantabria, CSIC-University of Cantabria, 39005 Santander, Spain}
\affiliation{Carnegie Mellon University, Pittsburgh, Pennsylvania 15213, USA}
\affiliation{Enrico Fermi Institute, University of Chicago, Chicago, Illinois 60637, USA}
\affiliation{Comenius University, 842 48 Bratislava, Slovakia; Institute of Experimental Physics, 040 01 Kosice, Slovakia}
\affiliation{Joint Institute for Nuclear Research, RU-141980 Dubna, Russia}
\affiliation{Duke University, Durham, North Carolina 27708, USA}
\affiliation{Fermi National Accelerator Laboratory, Batavia, Illinois 60510, USA}
\affiliation{University of Florida, Gainesville, Florida 32611, USA}
\affiliation{Laboratori Nazionali di Frascati, Istituto Nazionale di Fisica Nucleare, I-00044 Frascati, Italy}
\affiliation{University of Geneva, CH-1211 Geneva 4, Switzerland}
\affiliation{Glasgow University, Glasgow G12 8QQ, United Kingdom}
\affiliation{Harvard University, Cambridge, Massachusetts 02138, USA}
\affiliation{Division of High Energy Physics, Department of Physics, University of Helsinki and Helsinki Institute of Physics, FIN-00014, Helsinki, Finland}
\affiliation{University of Illinois, Urbana, Illinois 61801, USA}
\affiliation{The Johns Hopkins University, Baltimore, Maryland 21218, USA}
\affiliation{Institut f\"{u}r Experimentelle Kernphysik, Karlsruhe Institute of Technology, D-76131 Karlsruhe, Germany}
\affiliation{Center for High Energy Physics: Kyungpook National University, Daegu 702-701, Korea; Seoul National University, Seoul 151-742, Korea; Sungkyunkwan University, Suwon 440-746, Korea; Korea Institute of Science and Technology Information, Daejeon 305-806, Korea; Chonnam National University, Gwangju 500-757, Korea; Chonbuk National University, Jeonju 561-756, Korea}
\affiliation{Ernest Orlando Lawrence Berkeley National Laboratory, Berkeley, California 94720, USA}
\affiliation{University of Liverpool, Liverpool L69 7ZE, United Kingdom}
\affiliation{University College London, London WC1E 6BT, United Kingdom}
\affiliation{Centro de Investigaciones Energeticas Medioambientales y Tecnologicas, E-28040 Madrid, Spain}
\affiliation{Massachusetts Institute of Technology, Cambridge, Massachusetts 02139, USA}
\affiliation{Institute of Particle Physics: McGill University, Montr\'{e}al, Qu\'{e}bec, Canada H3A~2T8; Simon Fraser University, Burnaby, British Columbia, Canada V5A~1S6; University of Toronto, Toronto, Ontario, Canada M5S~1A7; and TRIUMF, Vancouver, British Columbia, Canada V6T~2A3}
\affiliation{University of Michigan, Ann Arbor, Michigan 48109, USA}
\affiliation{Michigan State University, East Lansing, Michigan 48824, USA}
\affiliation{Institution for Theoretical and Experimental Physics, ITEP, Moscow 117259, Russia}
\affiliation{University of New Mexico, Albuquerque, New Mexico 87131, USA}
\affiliation{The Ohio State University, Columbus, Ohio 43210, USA}
\affiliation{Okayama University, Okayama 700-8530, Japan}
\affiliation{Osaka City University, Osaka 588, Japan}
\affiliation{University of Oxford, Oxford OX1 3RH, United Kingdom}
\affiliation{Istituto Nazionale di Fisica Nucleare, Sezione di Padova-Trento, $^{ff}$University of Padova, I-35131 Padova, Italy}
\affiliation{University of Pennsylvania, Philadelphia, Pennsylvania 19104, USA}
\affiliation{Istituto Nazionale di Fisica Nucleare Pisa, $^{gg}$University of Pisa, $^{hh}$University of Siena and $^{ii}$Scuola Normale Superiore, I-56127 Pisa, Italy}
\affiliation{University of Pittsburgh, Pittsburgh, Pennsylvania 15260, USA}
\affiliation{Purdue University, West Lafayette, Indiana 47907, USA}
\affiliation{University of Rochester, Rochester, New York 14627, USA}
\affiliation{The Rockefeller University, New York, New York 10065, USA}
\affiliation{Istituto Nazionale di Fisica Nucleare, Sezione di Roma 1, $^{jj}$Sapienza Universit\`{a} di Roma, I-00185 Roma, Italy}
\affiliation{Rutgers University, Piscataway, New Jersey 08855, USA}
\affiliation{Texas A\&M University, College Station, Texas 77843, USA}
\affiliation{Istituto Nazionale di Fisica Nucleare Trieste/Udine, I-34100 Trieste, $^{kk}$University of Udine, I-33100 Udine, Italy}
\affiliation{University of Tsukuba, Tsukuba, Ibaraki 305, Japan}
\affiliation{Tufts University, Medford, Massachusetts 02155, USA}
\affiliation{University of Virginia, Charlottesville, Virginia 22906, USA}
\affiliation{Waseda University, Tokyo 169, Japan}
\affiliation{Wayne State University, Detroit, Michigan 48201, USA}
\affiliation{University of Wisconsin, Madison, Wisconsin 53706, USA}
\affiliation{Yale University, New Haven, Connecticut 06520, USA}

\author{T.~Aaltonen}
\affiliation{Division of High Energy Physics, Department of Physics, University of Helsinki and Helsinki Institute of Physics, FIN-00014, Helsinki, Finland}
\author{B.~\'{A}lvarez~Gonz\'{a}lez$^z$}
\affiliation{Instituto de Fisica de Cantabria, CSIC-University of Cantabria, 39005 Santander, Spain}
\author{S.~Amerio}
\affiliation{Istituto Nazionale di Fisica Nucleare, Sezione di Padova-Trento, $^{ff}$University of Padova, I-35131 Padova, Italy}
\author{D.~Amidei}
\affiliation{University of Michigan, Ann Arbor, Michigan 48109, USA}
\author{A.~Anastassov$^x$}
\affiliation{Fermi National Accelerator Laboratory, Batavia, Illinois 60510, USA}
\author{A.~Annovi}
\affiliation{Laboratori Nazionali di Frascati, Istituto Nazionale di Fisica Nucleare, I-00044 Frascati, Italy}
\author{J.~Antos}
\affiliation{Comenius University, 842 48 Bratislava, Slovakia; Institute of Experimental Physics, 040 01 Kosice, Slovakia}
\author{G.~Apollinari}
\affiliation{Fermi National Accelerator Laboratory, Batavia, Illinois 60510, USA}
\author{J.A.~Appel}
\affiliation{Fermi National Accelerator Laboratory, Batavia, Illinois 60510, USA}
\author{T.~Arisawa}
\affiliation{Waseda University, Tokyo 169, Japan}
% ...

\collaboration{CDF Collaboration\footnote[2]{With visitors from
$^a$Istituto Nazionale di Fisica Nucleare, Sezione di Cagliari, 09042 Monserrato (Cagliari), Italy,
$^b$University of California Irvine, Irvine, CA 92697, USA,
$^c$University of California Santa Barbara, Santa Barbara, CA 93106, USA,
$^d$University of California Santa Cruz, Santa Cruz, CA 95064, USA,
$^e$Institute of Physics, Academy of Sciences of the Czech Republic, Czech Republic,
$^f$CERN, CH-1211 Geneva, Switzerland,
$^g$Cornell University, Ithaca, NY 14853, USA,
$^h$University of Cyprus, Nicosia CY-1678, Cyprus,
$^i$Office of Science, U.S. Department of Energy, Washington, DC 20585, USA,
$^j$University College Dublin, Dublin 4, Ireland,
$^k$ETH, 8092 Zurich, Switzerland,
$^l$University of Fukui, Fukui City, Fukui Prefecture, Japan 910-0017,
$^m$Universidad Iberoamericana, Mexico D.F., Mexico,
$^n$University of Iowa, Iowa City, IA 52242, USA,
$^o$Kinki University, Higashi-Osaka City, Japan 577-8502,
$^p$Kansas State University, Manhattan, KS 66506, USA,
$^q$Ewha Womans University, Seoul, 120-750, Korea,
$^r$University of Manchester, Manchester M13 9PL, United Kingdom,
$^s$Queen Mary, University of London, London, E1 4NS, United Kingdom,
$^t$University of Melbourne, Victoria 3010, Australia,
$^u$Muons, Inc., Batavia, IL 60510, USA,
$^v$Nagasaki Institute of Applied Science, Nagasaki, Japan,
$^w$National Research Nuclear University, Moscow, Russia,
$^x$Northwestern University, Evanston, IL 60208, USA,
$^y$University of Notre Dame, Notre Dame, IN 46556, USA,
$^z$Universidad de Oviedo, E-33007 Oviedo, Spain,
$^{aa}$CNRS-IN2P3, Paris, F-75205 France,
$^{bb}$Texas Tech University, Lubbock, TX 79609, USA,
$^{cc}$Universidad Tecnica Federico Santa Maria, 110v Valparaiso, Chile,
$^{dd}$Yarmouk University, Irbid 211-63, Jordan.
}}
\noaffiliation
 
}

\ifthenelse{\equal{\AuthorList}{false}}{
\begin{figure}  
  \leftline{\includegraphics[scale=0.5]{cdfii_thumb_logo.eps}\hfill
    CDF/PUB/EXOTIC/CDFR/10876}
\end{figure}

\vspace*{0.5cm} 

\author{The CDF Collaboration}
\affiliation{URL http://www-cdf.fnal.gov}
}

\date{\today}

\begin{abstract}

\ifthenelse{\equal{\AuthorList}{false}}{
\vspace*{0.2cm}
}

We present a search for the standard model Higgs boson produced in
association with a $W$ boson in $\sqrt{s}=1.96$~TeV p-pbar collision
data collected with the CDF~II detector at the Tevatron corresponding
to an integrated luminosity of $9.45~\mathrm{fb^{-1}}$.  In events
consistent with the decay of the Higgs boson to a bottom-quark pair
and the $W$ boson to an electron or muon and a neutrino, we set $95\%$
credibility level upper limits on the $WH$ production cross section
times the $H\rightarrow b \bar{b}$ branching ratio as a function of
Higgs boson mass.  At a Higgs boson mass of $125$~\gevcc{} we observe
(expect) a limit of $4.9$ ($2.8$) times the standard model value.

\end{abstract}


\maketitle

%% Introduction
The mechanism of electroweak symmetry breaking~\cite{higgs} in the standard model (SM)~\cite{sm} predicts the existence of a fundamental scalar boson, the Higgs boson.  The SM does not predict the mass of the Higgs boson, ($m_H$), but through the combination of precision electroweak measurements~\cite{elweak}, including recent top quark and $W$ boson mass measurements from the Tevatron~\cite{top_quark_mass, w_boson_mass}, $m_H$ is constrained to be less than 152~\gevcc{} at the 95\% confidence level.  Direct searches at LEP2~\cite{sm-lep}, the Tevatron~\cite{tevcomb2012}, and the LHC~\cite{cmscomb2012,atlascomb2012} exclude possible masses of the SM Higgs boson at the 95\% confidence level or the 95\% credibility level (C.L.), except within the ranges 116.6 -- 119.4~\gevcc{} and 122.1 -- 127~\gevcc{}.  At the LHC experiments, sensitivity to the Higgs boson primarily comes from channels where the Higgs boson decays into two $W$ bosons, two photons, or two $Z$ bosons.  At the Tevatron, searches for a 116--127~\gevcc{} Higgs boson are most sensitive to the \bbbar{} final state, which offer the complementary information of the quark Yukawa couplings to the Higgs boson.  These searches may then be able to establish the mechanism of electroweak symmetry breaking as the source of fermionic mass in the quark sector.  

In the \bbbar{} final state, each $b$ quark fragments into a jet of hadrons and the Higgs boson signal can be reconstructed as an enhancement in the invariant mass distribution of these jets.  For a pair of jets, the dijet mass resolution at CDF is expected to be 10--15\% of the pair's mean reconstructed mass~\cite{Aaltonen:2011di}, which is approximately ten times larger than the reconstructed mass resolution in the leptonic or photonic search channels at the LHC.  Searches for the Higgs boson produced in association with a $W$ boson ($WH$), where the $W$ boson decays into a charged lepton ($\ell$) and a neutrino ($\nu$), provide the most sensitive search channel at the Tevatron in the mass range 116--127 \gevcc{}, because the requirements of a charged lepton candidate and of large missing transverse energy ($\MET$)~\cite{MET}, consistent with a neutrino escaping detection, significantly reduce the backgrounds from multijet processes.  Searches for the SM Higgs boson including this final state have been reported by the CDF, D0, ATLAS, and CMS collaborations~\cite{Aaltonen:2009dh,Abazov:2010hn,CDF:2011ae, cmsbb, atlasbb}.

%% Experiment and Trigger
In this Letter we describe a search for the Higgs boson in the \WHlvbb{} channel using the full data set collected during Run II of the Collider Detector experiment at the Fermilab Tevatron (CDF).  The CDF experiment is a general purpose detector described in Ref.~\cite{Acosta:2004yw}.  These data correspond to a luminosity of 9.45~\fb{} of \ppbar{} collisions.  Many aspects of the analysis remain unchanged from a recent search based on 7.5~\fb{} and are described in more detail in Ref.~\cite{Aaltonen:2012tf}.  Events are collected with on-line selection criteria (triggers) that require one of the following signatures: an electron candidate with transverse energy exceeding 18~\gevc{}~\cite{geometry}; a muon candidate with transverse momentum ($\pt$) exceeding $18$~\gevc{}; or $\METVEC(\mbox{cal}) > 15$~GeV with a forward ($|\eta|>1.2$) electromagnetic energy cluster satisfying $\et > 20$~GeV (designed to accept forward electrons from the $W$ boson decay).  An additional set of triggers is included that does not explicitly require an identified lepton, but instead requires $\METVEC(\mbox{cal}) > 45$~GeV or $\METVEC(\mbox{cal})>35$~GeV and a pair of jets~\cite{Buzatu:NIM}. 

%% Event selection
The identification of leptons and jets closely follows that for the CDF single-top-quark discovery described in Ref.~\cite{Aaltonen:2009jj}.  Candidate events are selected by requiring the presence of exactly one lepton candidate with $\pt>20$~\gevc.  The required $\MET$ is specific to each class of reconstructed lepton candidate to satisfy trigger requirements and suppress instrumental backgrounds; events with an electron satisfying $|\eta| \leq 1.1$, electron satisfying $|\eta| > 1.1$, non-isolated electron~\cite{isolation}, muon, or isolated track are required to have ~$\MET>20, 25, 25, 10$, or $20$~GeV, respectively.  Events are required to have exactly two or three jets satisfying $|\eta|<2.0$ and $E_T>20$~GeV after corrections for instrumental effects~\cite{Bhatti:2005ai}.  Events are rejected if they are kinematically inconsistent with leptonic $W$ boson decays as determined by a support vector model specific to each lepton category~\cite{Sforza:2011zz}.  Each support vector model is a binary classifier resulting from supervised training using information about the energies and angles of the lepton, jets and missing energy.

At least one of the jets must be identified (tagged) as consistent with the fragmentation of a $b$ quark according to a neural network tagging algorithm ~\cite{Freeman:2012uf}.  For each jet containing at least one charged particle track, the algorithm produces a scalar value in the range --1 to 1.   By comparing this value to two predetermined thresholds, the jet is classified as not tagged, \emph{loose tagged} (L), or \emph{tight tagged} (T), with all tight-tagged jets also satisfying the loose-tag definition. The thresholds are chosen to optimize the combined expected exclusion sensitivity in simulated events and the performance of the T and L $b$ tag selection is described in Ref.~\cite{Freeman:2012uf}.  The search sample is composed of seven orthogonal categories according to the exact number and type of $b$ tags in the event: TT, TL, T, LL, L for two-jet events, and TT, TL for three-jet events.  If an event satisfies two categories, the category of highest signal purity is chosen.  The inclusion of additional $b$-tag categories for events with three jets offers negligible improvement to the expected sensitivity and they are therefore not included.  The tagging algorithm and strategy employed here is identical to that described in the Tevatron combined observation of diboson production with decays to heavy-flavor quarks~\cite{TEVNPH:2012ab}.

%The definition of T (L) results in a per-jet tag rate of 42\% (70\%) for jets containing the fragmentation of a $b$ quark, 9\% (27\%) for jets containing the fragmentation of a charm quark and no $b$ quark, and 0.89\% (8.9\%) for jets without the fragmentation of a $b$ or charm quark.  

%Signal and BG model
 The Higgs boson events are modeled with the \pythia~\cite{pythia} Monte Carlo event generator combined with a detailed simulation of
  the CDF II detector~\cite{GEANT3,GFLASH} and tuned to the Tevatron underlying-event data~\cite{Aaltonen:2010rm}.   Small corrections to the simulated response of the detector are made based on data-simulation comparisons from orthogonal data sets~\cite{CDF:2011ae, Ketchum:2012zz}.  Models for background processes are derived from a mixture of simulation and data-driven techniques~\cite{Aaltonen:2010jr}.  Background processes to \WHlvbb{} include $W$ or $Z$ bosons produced in association with jets.  These processes may include true $b$ jets as in $W+b\bar{b}$, or non-$b$ jets that have been misidentified as $b$ jets like $W+c\bar{c}$, $W+cj$,  and $W+jj$, where $j$ refers to jets not originating from heavy-flavor quarks. Events with a top quark ($t{\bar{t}}$ and single-top-quark production), diboson events, and multijet events without $W$ bosons also contribute to the sample composition.

\begin{figure*}[th]
\begin{center}
\includegraphics[width=0.85\textwidth]{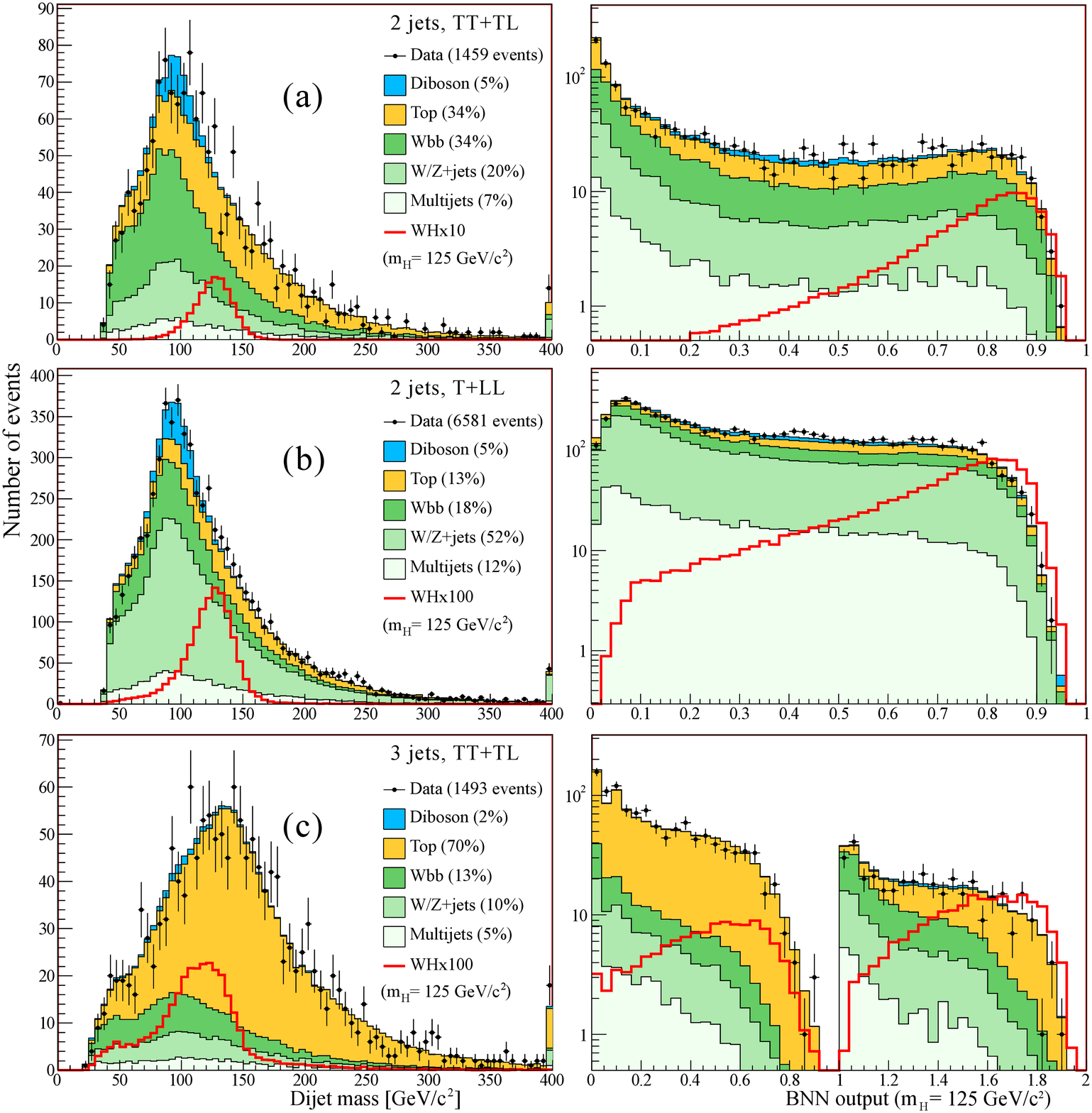}
\caption[]{The distribution for the dijet mass used as an input to the \BNN{} (left), and the \BNN{} output distribution (right).  Event $b$-tag categories of comparable signal purity are combined and presented as three orthogonal subsamples: two-jet TT+TL(a), two-jet T+LL(b), and three-jet TT+TL(c). The background is normalized to its prediction and the signal expectation of a Higgs boson mass of 125~\gevcc{} is scaled to 10, 100, and 100 times the SM prediction in (a), (b), and (c), respectively.  The right panel of (c) shows the \BNN{} distribution for events with exactly three jets, split into two regions based on an independent discriminant designed to separate top-quark-like events (assigned values between zero and one) from $W+$jets-like events (assigned values between one and two).  Statistical uncertainties are shown for the data points.}
\label{fig:stack}
\end{center}
\end{figure*}

The distributions of the reconstructed dijet invariant mass~\cite{multijet_mass, frank_thesis} of background and simulated Higgs boson events in the categories that contribute most to the sensitivity are shown in Fig. \ref{fig:stack}, with categories of comparable signal purity summed together.   The two-jet single-loose-tagged sample, L, contains twice as many events and has ten times smaller signal purity than the other two-jet categories combined.  This category contributes less than 1\% to the total expected exclusion sensitivity and is not presented in Fig. \ref{fig:stack}.  Event yields are stated as sums of categories corresponding to those presented in the figure.  The total expected signal yield in the current data set, assuming $m_H = 125$~\gevcc\ and based on next-to-next-to-leading-order (NNLO) theory predictions for the production rate~\cite{SMVHBR}, is $12.9 \pm 1.1$ $(12.4 \pm 0.9)$ for the TT+TL (T+LL) categories.  Events with exactly three jets account for $\sim$10\% of the total expected signal yield.  The background expectation of $1500 \pm 400$ ($6600 \pm 1900$) for TT+TL (T+LL) events is significantly larger than the expected number of signal events. The invariant mass of jets is the most discriminating signal variable between signal and background, but greater signal significance is achieved by using additional kinematic information available in each event.

%% Stack plots

%% NN
We employ a Bayesian artificial neural network (\BNN)~\cite{Peterson:1993nk} trained to discriminate \WHlvbb{} signal from the background using the
information contained in the following kinematic variables: 
the invariant mass of the candidate Higgs-boson-decay jets~\cite{multijet_mass};
the maximum invariant mass of the lepton, \METVEC, and one of the two jets ($\max (M_{\ell,\slashed E_T,j_1},M_{\ell,\slashed E_T,j_2})$); 
the lepton electric charge times its pseudorapidity;
the scalar sum of the lepton and jet transverse momenta minus the \MET, ($\sum_{\mbox{\tiny jets}}E_T+{\pt}^{\ell}-\slashed E_T$); 
the scalar sum of the transverse energy of calorimeter jets that fail the jet energy selection criteria, ($\sum_{\mbox{\tiny low-$E_T$ jets}}E_T$);
the absolute value of the transverse momentum of the reconstructed $W$ boson, reconstructed as $\pt^{\ell}+\METVEC$;
and the scalar sum of the jet, lepton and neutrino transverse energies, ($\sum_{\mbox{\tiny jets}}E_T+\pt^{\ell}+$\MET ).
The \BNN{} combines the discriminating power of these variables into a single output variable which, when used in searches for a 125~\gevcc{} Higgs boson, is capable of excluding cross sections times branching ratios 27\% lower in the background-only hypothesis as compared to searches using the jet invariant mass alone.  Improvements for other mass hypotheses are comparable.   We validate the predictions of the background model for each input variable in data control regions and we optimize the discriminants separately for each Higgs boson mass hypothesis. 
The distributions of the \BNN{} outputs of the neural network trained for a Higgs boson mass of 125~\gevcc{} are shown in the right panels of Fig.~\ref{fig:stack}.  Additional sensitivity from the three-jet categories is gained by training and employing a \BNN{} to separate top-quark-like from $W$+jets-like events, independently from the \BNN{} trained to separate $WH$ events from background.  In the right panel of Fig.~\ref{fig:stack}(c), top-quark-like events occupy the range of 0--1 of the discriminant, while $W$+jets-like events occupy the range 1--2. 

%% Statistical treatment
%Observing no evidence for Higgs boson production in the \BNN{} distribution, 

We calculate a Bayesian C.L. limit for each mass hypothesis using the combined binned likelihood of the \BNN{} output distributions.  Each of the seven jet-tagging categories are subdivided into four orthogonal lepton categories, depending on their distinct instrumental backgrounds.  After exclusion of two low-signal combinations~\cite{looseelectron3jet}, the analysis comprises 26 independent channels that are included in the likelihood.  The benefit of this subdivision of the search sample is both higher signal significance, and the isolation of individual background components for systematic constraint.  A posterior density is obtained by  multiplying this likelihood by Gaussian prior densities for the background normalizations and systematic uncertainties leaving $\sigma\times$\BRhbb{} with a uniform prior density, with priors truncated to prevent negative predictions.  A 95\% C.L. limit is determined such that 95\% of the posterior density for $\sigma\times$\BRhbb{} accumulates below the limit~\cite{pdg}.   

Systematic uncertainties on the rate of signal and background production from jet energy scale, $b$-tagging efficiencies, lepton identification and trigger efficiencies, the amount of initial and final state radiation (ISR and FSR), and the parton distribution functions are included in the limit calculation~\cite{Buzatu:2011zz}.  In addition, the limit calculation includes shape uncertainties on the discriminant output~\cite{Aaltonen:2011rt}, arising from uncertainties on the jet energy scale, ISR and FSR for all simulated samples, and arising from uncertainties on the renormalization and factorization scale for $W$+jets samples.  The expected exclusion limits are $\sim$20\% tighter if the calculation is performed without including systematic uncertainties.  The impact of kinematic differences between simulated and data events $W$+jets is investigated as a potential source of systematic uncertainty.  The jet energies, angular separations, and invariant mass distributions of events selected prior to $b$ tagging are used to derive shape corrections which are applied to simulated $W$+jets events in the search samples.  These adjustments show negligible impact on the discriminant shape of the background prediction, and therefore are not considered in the final results.

\begin{figure}
\begin{center}
\includegraphics[width=\columnwidth]{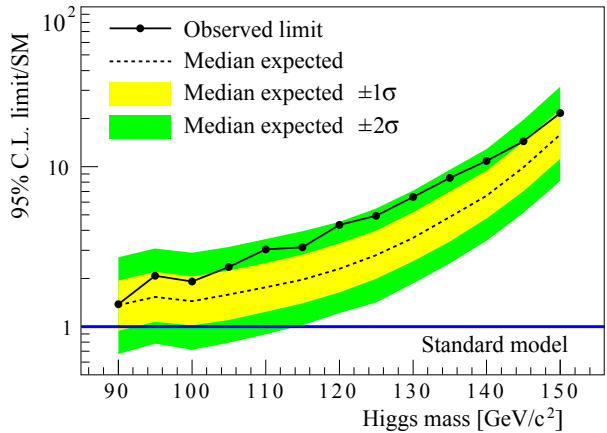}
\caption[]{
The observed 95\% C.L. upper limits on Higgs boson production relative to 
the SM expectation as a function of the Higgs boson mass.  The median 
expected limits in the absence of a Higgs signal are indicated by the  
dashed line and the light and dark bands indicate the one and two sigma 
ranges for individual experiments in this background-only scenario.}
\label{fig:limits}
\end{center}
\end{figure}

%% Results
  Table~\ref{tab:limits} and Fig.~\ref{fig:limits} show the expected and observed limits calculated for different Higgs boson masses.  We find an observed (expected) 95\% C.L. limit of 4.9 (2.8) times the SM prediction of the production cross section times branching fraction for a Higgs boson mass of 125~\gevcc{} (NNLO theory predicts $\sigma\times$\BRhbb$=75~\mbox{fb}$)~\cite{SMVHBR}. The resulting expected exclusion limit is approximately a factor of 2.6 lower than our previous Letter~\cite{Aaltonen:2009dh}.  This improvement in expected sensitivity consists of a factor of  approximately 1.9 due to the increased data set~\cite{lumi_note} and a factor of approximately 1.4 due to analysis technique improvements.  Increased signal acceptance and background rejection gained from the improved $b$-tagging algorithm provide approximately 11\% improvement in exclusion sensitivity.  The inclusion of three-jet events, increased trigger acceptance, improved rejection of multijet events, and additional lepton acceptance via new reconstruction categories dominate the remaining improvement.  The two-jet TT and TL categories offer the highest signal purity, driving the sensitivity of the analysis.  Performing the analysis using these two categories alone produces expected and observed limits comparable to the full analysis combination, with an observed (expected) limit of 4.8 (3.2) times the SM $\sigma\times$\BRhbb{} for $m_H = 125$~\gevcc.  The consistency of the observed limits with the signal hypothesis is tested by statistical sampling of the signal-plus-background model. These studies indicate that the median upper C.L. in the SM Higgs scenario is $\sim$1 unit of SM cross-section higher than that for the background-only hypothesis over most of the 90--150~\gevcc{} search range, which is consistent with the observed limits to within one standard deviation.

%Conclusion and outlook

  In conclusion, we have presented a search for the SM Higgs boson produced in association with a $W$ boson using the complete CDF Run~II data set.  This analysis employs methods used in CDF analyses of well established SM processes, providing confidence in the robustness of the background model and search techniques.  The observed exclusion limits exceed those expected in the background-only scenario over much of the 90--150~\gevcc{} search range, with deviations from the background-only-hypothesis corresponding to local significances for tested Higgs boson masses between 120 and 135~\gevcc of roughly two sigma.  While the LHC experiments have surpassed the Tevatron experiments in overall sensitivity to a SM Higgs boson, the \WHlvbb{} search reported here is currently the most sensitive single-channel search for a low-mass SM Higgs boson in its favored decay mode.

%% Limit Table
\begin{table*}[tbh]
\begin{center}
\caption[]{
The SM prediction for $\sigma\times$\BRhbb, as well as the
expected and observed limits at 95\% C.L. on the Higgs boson
production cross section divided by the SM prediction as shown in
Fig.~\ref{fig:limits}.} 
\begin{tabular}{l c c c c c c c c c c c c c}\hline\hline
Mass (\gevcc) &   90  &  95 & 100  & 105  & 110  & 115  & 120  & 125  & 130  & 135  & 140
& 145  & 150  \\ \hline
SM $\sigma\times$\BRhbb{} (fb) & 321&268 & 223 & 185 & 152 & 123 & 98  & 75 & 55 & 39 & 27 &  17& 10 \\ \hline
Exp. (95\% C.L./SM) & 1.36 & 1.53 & 1.44 & 1.58 & 1.76 & 1.97 & 2.30 & 2.79 & 3.59 & 4.85 & 6.59 & 9.91 & 15.9 \\
Observed  (95\% C.L./SM) & 1.38 & 2.07 & 1.92 & 2.36 & 3.03 & 3.13 & 4.33 & 4.93 & 6.47 & 8.51 & 10.9 & 14.4 & 21.7 \\
\hline\hline
\end{tabular}
\label{tab:limits}
\end{center}
\end{table*}

%% Limit figure

We thank the Fermilab staff and the technical staffs of the
participating institutions for their vital contributions. This work
was supported by the U.S. Department of Energy and National Science
Foundation; the Italian Istituto Nazionale di Fisica Nucleare; the
Ministry of Education, Culture, Sports, Science and Technology of
Japan; the Natural Sciences and Engineering Research Council of
Canada; the National Science Council of the Republic of China; the
Swiss National Science Foundation; the A.P. Sloan Foundation; the
Bundesministerium f\"ur Bildung und Forschung, Germany; the Korean
World Class University Program, the National Research Foundation of
Korea; the Science and Technology Facilities Council and the Royal
Society, UK; the Russian Foundation for Basic Research; the Ministerio
de Ciencia e Innovaci\'{o}n, and Programa Consolider-Ingenio 2010,
Spain; the Slovak R\&D Agency; the Academy of Finland; and the
Australian Research Council (ARC).

%------------------------------------------------------------------------------%       Bibliography
\bibliographystyle{apsrev}

%\bibliography{WHPRL2012}

\begin{thebibliography}{40}
\bibitem[{MET()}]{MET}
\bibinfo{note}{The calorimeter missing $E_T$ ($\METVEC(\mbox{cal})$) is defined
  by the sum over calorimeter towers, $\METVEC(\mbox{cal}) = - \sum_{i} E_T^i
  \hat{n}_i$, where $i$ is calorimeter tower number with $|\eta| < 3.6$,
  $\hat{n}_i$ is a unit vector perpendicular to the beam axis and pointing at
  the $i$th calorimeter tower. The reconstructed missing energy, $\METVEC$, is
  derived by subtracting from $\METVEC(\mbox{cal})$ components of the event not
  registered by the calorimeter, such as muons and jet energy adjustments.
  $\MET(\mbox{cal})$ and $\MET$ are the scalar magnitudes of
  $\METVEC(\mbox{cal})$ and $\METVEC$, respectively.}
\bibitem[{geo()}]{geometry}
\bibinfo{note}{We use a cylindrical coordinate system with the origin at the
  center of the CDF detector, $z$ pointing in the direction of the proton beam,
  $\theta$ and $\phi$ representing the polar and azimuthal angles,
  respectively, and pseudorapidity defined by $\eta= - \rm ln$ $\rm
  tan(\theta/2)$. The transverse momentum $p_T$ (transverse energy $E_{T}$) is
  defined to be $p\sin\theta$ ($E\sin\theta$).}
\bibitem[{mul()}]{multijet_mass}
\bibinfo{note}{The reconstructed mass of the Higgs boson candidate is chosen as
  the dijet pair in the two-jet categories, and is reconstructed by two methods
  in the three-jet category. One method chooses the two tagged jets, while the
  other method chooses two or three jets, depending on the kinematic properties
  of the event.}
\bibitem[{loo()}]{looseelectron3jet}
\bibinfo{note}{We identify three classes of electron candidates, two classes of
  triggered muons, and one class of muons taken via missing energy and jets
  triggers. In the limit calculation, the highest purity electron sample was
  combined with the triggered muon samples, yielding four lepton categories.
  The lowest purity category of electrons was not used in the three-jet
  analysis.}
\bibitem[{lum()}]{lumi_note}
\bibinfo{note}{We assume that the sensitivity would scale inversely to the
  square root of the integrated luminosity in the absence of analysis
  improvements.}
\end{thebibliography}

%''Higgs'' papers
%%CITATION = PRLTA,13,508;%%
%%CITATION = PRLTA,13,321;%%
%%CITATION = PRLTA,13,585;%%

%Top Mass combo
%%CITATION = ARXIV:1107.5255;%%

%EWK constraints
%%CITATION = ARXIV:1012.2367;%%

%W Mass
%%CITATION = ARXIV:1204.0042;%%

%LEP limits
%%CITATION = PHLTA,B565,61;%%

%Tevatron Higgs
%%CITATION = ARXIV:1207.0449;%%

% LHC COMBOs
%%CITATION = ARXIV:1202.1488;%%
%%CITATION = ARXIV:1207.0319;%%

%vh theory
%%CITATION = ARXIV:1003.4266;%%
%%CITATION = HEP-PH/0307206;%%

%vh theory
%%CITATION = ARXIV:1003.4266;%%
%%CITATION = HEP-PH/0307206;%%

%strange
%%CITATION = HEP-PH/9404247;%%
%%CITATION = HEP-PH/9309294;%%

%Dittmaier
%%CITATION = ARXIV:1201.3084;%%

%triggers
%%CITATION = ARXIV:1206.4813;%%

%Taggers
%%CITATION = ARXIV:1205.1812;%%

%%CITATION = ARXIV:1203.3782;%%

%PYTHIA
%%CITATION = HEP-PH/0108264;%%

%CTEQ%
%%CITATION = EPHJA,C12,375;%%

%HDECAY
%%CITATION = HEP-PH/9704448;%%

%UE
%%CITATION = ARXIV:1003.3146;%%

%single top
%%CITATION = ARXIV:0903.0885;%%

%past WH
%%CITATION = ARXIV:0906.5613;%%

%Tev Diboson
%%CITATION = ARXIV:1203.3782;%%
\end{document}